\documentstyle[12pt]{article}
\begin{document}
\begin{center}{\large{\bf Mapping and Embedding  of Two Metrics  Associated  with Dark Matter, Dark Energy, and Ordinary Matter}}
\end{center}
\vspace*{1.5cm}
\begin{center}
A. C. V. V. de Siqueira
$^{*}$ \\
Departamento de Educa\c{c}\~ao\\
Universidade Federal Rural de Pernambuco \\
52.171-900, Recife, PE, Brazil.\\
\end{center}
\vspace*{1.5cm}
\begin{center}{\bf Abstract}

In this paper we build a mapping between two different metrics  and
embed them in a flat manifold. One of the metrics represents the
ordinary matter, and the other describes the dark matter, the dark
energy, and the particle-antiparticle asymmetry.The latter was
obtained in a recent paper. For the mapping and embedding we use two
new formalisms developed and presented in two previous papers,
\textbf{Mapping Among Manifolds} and, \textbf{Conformal Form of
Pseudo-Riemannian Metrics by Normal Coordinate Transformations},
which was a generalization of the Cartan's approach of Riemannian
normal coordinates.
\end{center}

 \vspace{3cm}

${}^*$ E-mail: acvvs@ded.ufrpe.br
\newline

\newpage

\section{Introduction}
$         $

The Einstein's theory of  general relativity is still the best
theory to describe problems in astrophysics and cosmology. However,
more recent observations in these two areas are apparently difficult
to be explained by general relativity. It raises the possibility to
consider models involving membranes and parallel universes, dark
matter, dark energy, and the cosmological constant to explain the
behavior of large scale structures like galaxies, clusters of
galaxies, and the universe. In a previous paper we obtained a
spatially flat solution of the Einstein's equation with a
Klein-Gordon field and the cosmological constant \cite{1}. We have
shown in a recent paper \cite{2}, that this scalar particle solution
is a candidate to explain the possible origin of dark matter, dark
energy, and particle-antiparticle asymmetry. The introduction of a
new field in the Einstein's equation could destroy our dark matter
and dark energy solution. If it is a good candidate to describe dark
matter, dark energy, and particle-antiparticle asymmetry, then the
Einstein's equation needs to be preserved  as in \cite{2}. The
embedding of one metric only  in an n-dimensional flat space  is
well known. The embedding of a classical metric and the dark matter
and dark energy metric in an n-dimensional flat space is a possible
strategy in order to consider the gravitational interaction between
the scalar particle and the ordinary matter.
\newline
This paper is organized as follows. In Sec. $2$, we present a
primordial and spatially flat solution of the Einstein's equation
with a massive scalar Klein-Gordon field and the cosmological
constant. The Jacobi equation is presented for this solution and two
primordial forces are identified as dark matter and dark energy,
respectively. In Sec. $3$, we present some of our results in
\textbf{Conformal Form of Pseudo-Riemannian Metrics by Normal
Coordinate Transformations} as a generalization of the Cartan's
approach of Riemannian normal coordinates \cite{3}. In Sec. $4$, we
present some of our results in \textbf{Mapping Among Manifolds
}\cite{4}. As we will see, these are important for the embedding of
two different manifolds in an n-dimensional flat manifold. In Sec.
$5$, we build a mapping between dark matter, dark energy and
ordinary matter. In Sec. $6$, we build an embedding of two metrics,
one associated with ordinary matter and the other with dark matter
and dark energy. In Sec. $7$, we present the two $6$-dimensional
hyper-vectors, normal to the dark matter-dark energy manifold, and
show how to build two $6$-dimensional hyper-vectors, normal to the
ordinary matter manifold, and how to make the mapping among
$6$-dimensional hyper-vectors. In Sec.$8$, we summarize and conclude
the results of this paper.

\renewcommand{\theequation}{\thesection.\arabic{equation}}
\section{\bf An Exact Solution of the Einstein's Equation}
\setcounter{equation}{0}
 $         $
In a previous paper we obtained three solutions of the Einstein's
equation with a  Klein-Gordon field and the cosmological constant
\cite{1}. In a recent paper we presented in details some aspects of
the spatially flat solution relevant to dark matter, dark energy,
and particle-antiparticle asymmetry \cite{2}. In this paper it will
be necessary to introduce  some of those aspects as a short review.
The convention used in a local basis was \cite{1}, \cite{2},
\begin{equation}
R^{\alpha}{}_{\mu \sigma \nu }=\partial_{\nu} \Gamma_{\mu \sigma
}^{\alpha}-\partial_{\sigma}\Gamma_{\mu \nu}^{\alpha}+\Gamma_{\mu
\sigma}^{\eta}\Gamma_{n \nu}^{\alpha}-\Gamma_{\mu \nu}^{\eta}\Gamma
_{\sigma \eta}^{\alpha}
\end{equation}
with  Ricci tensor
\begin{equation}
R_{\mu \nu}=R^{\alpha}{_{\mu \alpha \nu }}.
\end{equation}
For this convention we have the following Einstein`s equation, with
the cosmological constant $\Lambda $,
\begin{equation}
R_{\mu \nu }-\frac{1}{2}g_{\mu \nu }R+\Lambda g_{\mu
\nu}=-\frac{8\pi G}{ c^{2}}T_{\mu \nu}
\end{equation}
where $T_{\mu \nu}$ is the momentum-energy tensor of a massive
scalar field,
\begin{equation}
T_{\mu \nu }=2 \nabla_{\mu} \phi \nabla_{\nu} \phi -g_{\mu
\nu}\nabla^{\alpha } \phi \nabla_{\alpha}\phi +m^{2} g_{\mu \nu
}\phi^{2}
\end{equation}
We have used (+, -, -, -) signature convention and a
Friedmann-Robertson-Walker line element given by
\begin{equation}
ds^{2}=dt^{2}-\frac{d\sigma^{2}e^{g}}{\left( 1+Br^{2}\right)^{2}}
\end{equation}
where $d \sigma^{2}$ is the three-dimensional Euclidian line element
and $A=8\pi G/c^{2}$, $B=k/4a'^{2}$ and with $ k=0$ , $k=1$ and
$k=-1$. We also have $a'^{2}$ as a constant.
\newline
We pay attention to our spatial flat solution,  $B=0$. In this case
the field is given by
\begin{equation}
\phi =\frac{\in mt}{\sqrt{3A}}+b
\end{equation}
with $\in =\pm 1$ and $b$ as an arbitrary constant. The cosmological
constant obeys the condition
\begin{equation}
\Lambda =-\frac{m^{2}}{3}=-\frac{1}{3}( \frac{cM}{\hbar})^2,
\end{equation}
a negative value, associated with the Planck's constant, the speed
of light and a scalar particle of mass $M$.
\newline
The corresponding line element is
\begin{equation}
ds^{2}=dt^{2}-d\sigma ^{2}e^{[-2\in
mb(\sqrt{\frac{A}{3}})t-\frac{m^{2}}{3} t^{2}]}.
\end{equation}
In this paper we use another convention to the Riemann tensor, as
follows,
\begin{equation}
R^{\alpha}{}_{\mu \sigma \nu }=-\partial_{\nu} \Gamma_{\mu \sigma
}^{\alpha}+\partial_{\sigma}\Gamma_{\mu \nu}^{\alpha}-\Gamma_{\mu
\sigma}^{\eta}\Gamma_{n \nu}^{\alpha}+\Gamma_{\mu \nu}^{\eta}\Gamma
_{\sigma \eta}^{\alpha}
\end{equation}
which implies
\begin{equation}
R_{\mu \nu }-\frac{1}{2}g_{\mu \nu }R-\Lambda g_{\mu \nu}=\frac{8\pi
G}{ c^{2}}T_{\mu \nu}.
\end{equation}
The motion will be simpler in a Fermi-Walker transported tetrad
basis.
\newline
Let us consider the connection between the tetrad and the local
metric tensor
\begin{equation}
g_{\lambda\pi}=E_{\lambda}^{(\mathbf{A})}E_{\pi}^{(\mathbf{B})}\eta_{(\mathbf{A})(\mathbf{B})},
\end{equation}
where $ \eta_{(\mathbf{A})(\mathbf{B})}$ and $
E_{\lambda}^{(\mathbf{A})}$ are the Lorentzian metric and tetrad
components, respectively.
\newpage
From (2.8) we have
\begin{equation}
E_{0}^{(\mathbf{0})}=1,
\end{equation}
\begin{equation}
E_{1}^{(\mathbf{1})}=E_{2}^{(\mathbf{2})}=E_{3}^{(\mathbf{3})}=e^{[-\in
mb(\sqrt{\frac{A}{3}})t-\frac{m^{2}}{6} t^{2}]}.
\end{equation}
We now write the $1$-form
\begin{equation}
\theta^{(\mathbf{A})}= dx^{\lambda} E_{\lambda}^{(\mathbf{A})}.
\end{equation}
By exterior derivatives of (2.14) and using the Cartan's second
structure equation, we obtain
\begin{eqnarray}
 \nonumber R^{(\mathbf{1})}{_{(\mathbf{0})(\mathbf{0})(\mathbf{1})}}=R^{(\mathbf{2})}{_{(\mathbf{0})(\mathbf{0})(\mathbf{2})}}=\\
 \nonumber
 =R^{(\mathbf{3})}{_{(\mathbf{0})(\mathbf{0})(\mathbf{3})}}=-\frac{m^{2}}{3}+\frac{1}{2}[-2\in
mb(\sqrt{\frac{A}{3}})-\frac{m^{2}}{3}t]^{2}.\\
\end{eqnarray}
Let us present the Jacobi equation in a Fermi-Walker transported
tetrad basis \cite{6},
\begin{equation}
\frac{d^2Z^{(\mathbf{A})}}{dt
^2}+R^{(\mathbf{A})}{_{(\mathbf{0})(\mathbf{C})(\mathbf{0})}}Z^{(\mathbf{C})}=0.
\end{equation}
Substituting $(2.15)$ in $(2.16)$ we obtain
\begin{equation}
\frac{d^2Z^{(\mathbf{A})}}{dt
^2}=\{-\frac{m^{2}}{3}+\frac{1}{2}[-2\in
mb(\sqrt{\frac{A}{3}})-\frac{m^{2}}{3}t]^{2}\}Z^{(\mathbf{A})},
\end{equation}
with $A=(1,2,3)$.
\newline
We can rewrite $(2.17)$ as follows
\begin{equation}
\frac{d^2Z^{(\mathbf{A})}}{dt
^2}=\{-\frac{m^{2}}{3}-2bm^{3}\frac{1}{3}(\sqrt{\frac{A}{3}})t
+\frac{m^{2}}{3}b^{2}A+\frac{m^{4}}{18}t^{2}\}Z^{(\mathbf{A})}.
\end{equation}
The Jacobi equation will be appropriate to show the relative
acceleration between two particles  if we do not have to consider
the metric deformation by particles. For the primordial universe
(2.8), we have from $(2.17)$ or $(2.18)$ that two  massive scalar
particles in two geodesics close to each other feel two primordial
forces, being one attractive (dark matter) and the other repulsive
(dark energy), both increasing with distance. The same scalar
particle will be responsible for the two metric forces which,
conveniently, we have identified as dark matter and dark energy.
From the gravitational point of view, the creation of other types of
matter by the universe generates three competitive forces, the two
primordial forces above presented, and another which, for galaxies,
can be expressed by the Newtonian gravity. Inside and outside the
galaxies, the resulting force is the sum of these three forces. A
correct dynamic description of one or more stars in a galaxy,
depends on a set of information about galaxy evolution. Elliptical
and spiral galaxies, as well as clusters of galaxies, have different
dynamics and different evolution processes. The Newtonian gravity is
very important to describe the galaxies dynamics but it is not
enough. Physicists and astronomers have concluded that the Newtonian
gravity only is not sufficient to describe the galaxies dynamics.
They believe in the existence of a second attractive force (dark
matter) which, in association with the Newtonian gravity, governs
the star dynamics. They also believe in the existence of a repulsive
force (dark energy) responsible for the expansion on large scale. We
believe that the presence of the two primordial forces together with
the Newtonian force can describe the galaxies behavior. Inside and
outside the galaxies the resulting force is the sum of the three
forces. The value of the constant $b$ in (2.6) and (2.8) could be
fixed by experimental records of galaxies (dark matter) or
cosmological expansion (dark energy) or both. Modifications in the
stars motion in galaxies can be done by appropriate adjustments in
the constant $b$. It is possible  that $b$ is a new constant of
nature, as well as the mass $M$ of the scalar particle.
\newline
The mass of the scalar particle can be estimated by astronomic
measurements of the cosmological constant. From $(2.7)$ and $(2.17)$
we conclude that $\Lambda$ is associated with an attractive force,
for us conveniently identified as dark matter, and a repulsive
force, identified as dark energy. The second term in the second
member of $(2.17)$ is positive and, therefore, identified as dark
energy. Notice that the dark energy term is a function of time, of
the constant $b$, and also of the term identified as  dark matter.
We noticed in $(2.18)$ that the constants $m$ and $b$ are present in
the terms associated with the dark matter (attractive force) and to
the terms associated with the dark energy (repulsive force). Then,
$\Lambda$ is present in the dark matter and in the dark energy. In
other words, it is not possible to separate them, because dark
matter and dark energy are scalar particle effects. Originally, the
cosmological constant was associated with a repulsive force so that
the estimate given in the following is associated with a large scale
expansion. It can be slightly different from a realistic and
definitive value, but our objective is to point that we cannot
detect the scalar particle, or, at best, we have a very low
probability of doing so.
\newline
Using the experimental limit for the constant  $\Lambda$ in (2.7)
\cite{6}, it is possible to obtain  a superior limit for the mass
 of the scalar particle. The cosmological constant was estimated as
\begin{equation}
\Lambda<{10}^{-54}cm^{-2}.
\end{equation}
Using it and (2.7) we obtain
\begin{equation}
M<(6).{10}^{-65}g.
\end{equation}
There is another limit for the cosmological constant \cite{7} given
by
\begin{equation}
\Lambda{Lp}^2<{10}^{-123},
\end{equation}
or
\begin{equation}
\Lambda<{10}^{-57}cm^{-2}.
\end{equation}
Using it and (2.7) we obtain
\begin{equation}
M<(1.9).{10}^{-66}g.
\end{equation}
The relationship between the electron rest mass and the mass of the
scalar particle is approximately given by
\begin{equation}
m_{e}\sim (4.79).{10}^{38}M.
\end{equation}
The universe expansion can be calculated. In other words, it is
possible to calculate the starting point of the universe
contraction. Using (2.31) and (2.8) we obtain
\begin{equation}
t\sim\frac{1}{\sqrt{-\Lambda}}.
\end{equation}
Notice that we have assumed $c=1$ in (2.8).  Therefore, for numeric
results, involving time, we regain $ct$, so that
\begin{equation}
t\sim{10}^{27}\frac{cm}{c}\sim(3.33){10}^{9}years,
\end{equation}
where $(2.19)$ and $(2.25)$ were used.
\newline
Using $(2.22)$ in $(2.25)$ we have
\begin{equation}
t\sim(3.162){10}^{28}\frac{cm}{c}\sim(33.4){10}^{9}years.
\end{equation}
Note that $(2.26)$ is incompatible with the geological data of the
Earth. If $(2.27)$ is  a good estimate, it will be almost impossible
to detect the scalar particle. Its influence will be predominantly
gravitational and it is given by (2.8). Consequently, for many
classical situations as, for instance, the solar system dynamics,
the effect on the ordinary matter would be insignificant. For this
condition we consider only the ordinary matter in the Einstein's
equation. The scalar particle can be very important for galaxies,
clusters of galaxies, and large structures. It is necessary a
investigation to evaluate the influence of an intense gravitational
field generated by a classical black hole geometry on the primordial
scalar particle.
\newline
The primordial universe (2.8) starts with scalar particles and is
non-singular  at $t=0$. It is an expansible universe if the
curvature ${R^{(A)}}_{(A)}$ obeys a simple inequality. With the time
evolution, other types of matter were created and complex
interactions among particles are checked every day. Analytical
solutions of (2.10) with the inclusion of other fields are very
difficult. However, as the influence of the primordial universe
(2.8) could have been very important in the past and can be very
important in the present, it is reasonable to suppose that other
primordial particles are ghosts, so that (2.8) is a consequence of
the scalar particles only. In other words, the momentum-energy
tensors of other primordial fields have not contributed to the
curvature of the primordial universe in the past nor in the present,
although such particles interact with all that, playing an important
part in the evolution of the universe, as well as in the creation of
the ordinary matter. We recall that in the cosmological models,
metrics as the Friedmann-Robertson-Walker are important to the
initial large structure formation, as well as to the universe
evolution. But, gradually each local distribution of matter will be
more and more important and the effect of all distributions of
matter in the universe is represented by a momentum-energy tensor of
a fluid in the Einstein's equation for a Friedmann-Robertson-Walker
metric. However, if (2.8) is responsible for the dark matter and the
dark energy, we will have a different situation. In this case (2.8)
would determine the evolution of the universe in the past and in the
present and, due to the mass estimate of the scalar particle, its
interaction with other particles would be predominantly
gravitational.
\newline
It is important to notice the presence of two different time scales,
one associated with a local distribution, as well as with a large
scale structure of ordinary matter, and the other associated with
the cosmological time of (2.8). The embedding of (2.8) and of a
classical metric in an $n$ dimensional flat space is a possible
strategy to consider the gravitational interaction between the
scalar particles and the ordinary matter.
\newline
The primordial universe (2.8) is non-singular at $t=0$. It is
cyclical and eternal, and could have different cycles. Although it
is not the only possibility, a negative curvature is the simplest
mechanism for an expansive universe and it will be considered.
\newline
For (2.8) the curvature is given by
\begin{eqnarray}
 \nonumber{R^{(A)}}_{(A)} =\\
 \nonumber
 =-2(\frac{cM}{\hbar})^2 +6[-2\in
\frac{cM}{\hbar}b(\sqrt{\frac{A}{3}})-\frac{(\frac{cM}{\hbar})^2 }{3}t]^{2}.\\
\end{eqnarray}
We have at $t=0$
\begin{eqnarray}
 \nonumber{R^{(A)}}_{(A)}(t=0) =\\
 \nonumber
 =2(\frac{cM}{\hbar})^2[-1+4b^2.A],\\
\end{eqnarray}
which is a finite curvature. We consider an initial negative
curvature ${R^{(A)}}_{(A)}$ as the simplest condition for the
primordial expansive universe
\begin{eqnarray}
 \nonumber{R^{(A)}}_{(A)}(t=0) <0,\\
\end{eqnarray}
so that
\begin{eqnarray}
\| b\|<\frac{1}{2\sqrt{A}},
\end{eqnarray}
or
\begin{eqnarray}
\| b\|<\frac{c}{2\sqrt{8{\pi}G}},
\end{eqnarray}
or
\begin{eqnarray}
\| b\|<{10}^{15}\sqrt{\frac{g}{cm}},
\end{eqnarray}
where (2.33) is a superior limit for $b.$ We have obtained two
superior limits for the mass of the scalar particle and for the
constant $b$ given by (2.20) and (2.33), respectively. Note that $b$
and $M$ can be two new constants of nature. For (2.22), $M$ will be
given by (2.23), smaller than (2.20), reinforcing the previous
conclusion that it is very difficult to detect this scalar particle.
\newline
Note that our choice of an initial negative curvature, as the
expansion mechanism,  imposed a superior limit for the constant b.
However, other mechanisms are possible, so that the constant b can
assume another limit, compatible with experimental results.
\renewcommand{\theequation}{\thesection.\arabic{equation}}
\section{\bf Conformal Form of a Pseudo-Riemannian Metric by Normal Coordinate
Transformations}
 \setcounter{equation}{0}
 $         $
In a previous \cite{3} paper we extended the Cartan's approach of
Riemannian normal coordinates and showed that all n-dimensional
pseudo-Riemannian metrics are conformal to an  n-dimensional flat
manifold, as well as to an n-dimensional manifold of constant
curvature, when, in normal coordinates, they are well-behaved in the
origin and in its neighborhood. As a consequence of geometry,
without postulates, we obtained the classical and the quantum
angular momenta of a particle . In this Section a short review of
this approach will be presented.
\newline
Let us consider the line element
\begin{equation}
 ds^2= G_{\Lambda\Pi}du^{\Lambda}du^{\Pi},
\end{equation}
with
\begin{equation}
G_{\Lambda\Pi}=E_{\Lambda}^{(\mathbf{A})}E_{\Pi}^{(\mathbf{B})}\eta_{(\mathbf{A})(\mathbf{B})},
\end{equation}
where $ \eta_{(\mathbf{A})(\mathbf{B})}$ and $
E_{\Lambda}^{(\mathbf{A})}$ are flat metric and vielbein components,
respectively.
\newline
We choose each $ \eta_{(\mathbf{A})(\mathbf{B})}$ as plus or minus
Kronecker's  delta function, where a Lorentzian metric signature
will be a particular case.
\newline
Let us give the 1-form $\omega^{(\mathbf{A})} $ by
\begin{equation}
\omega^{(\mathbf{A})}= du^{\Lambda} E_{\Lambda}^{(\mathbf{A})}.
\end{equation}
We now define Riemannian normal coordinates by
\begin{equation}
  u^{\Lambda}=v^{\Lambda}t.
\end{equation}
Substituting in $(3.3)$
\begin{equation}
\omega^{(\mathbf{A})}= tdv^{\Lambda}
E_{\Lambda}^{(\mathbf{A})}+dtv^{\Lambda}E_{\Lambda}^{(\mathbf{A})}.
\end{equation}
Let us define
\begin{equation}
 z^{(\mathbf{A})}=v^{\Lambda}E_{\Lambda}^{(\mathbf{A})},
\end{equation}
so that
\begin{equation}
\omega^{(\mathbf{A})}=dtz^{(\mathbf{A})}+tdz^{(\mathbf{A})}
+tE^{\Pi(\mathbf{A})}\frac{\partial{E_{\Pi(\mathbf{B})}}}{\partial{z^{(\mathbf{C})}}}z^{(\mathbf{B})}dz^{(\mathbf{C})}.\\
\end{equation}
We now make
\begin{equation}
 A^{(\mathbf{A})_{(\mathbf{B})(\mathbf{C})}}=tE^{\Pi(\mathbf{A})}\frac{\partial{E_{\Pi(\mathbf{B})}}}{\partial{z^{(\mathbf{C})}}},
\end{equation}
then
\begin{equation}
\varpi^{(\mathbf{A})}=
tdz^{(\mathbf{A})}+A^{({A})_{(\mathbf{B})(\mathbf{C})}}z^{(\mathbf{B})}dz^{(\mathbf{C})},
\end{equation}
with
\begin{equation}
\omega^{(\mathbf{A})}=dtz^{(\mathbf{A})}+\varpi^{(\mathbf{A})}.
\end{equation}
We have at $t=0$
\begin{equation}
 A^{({A})_{(\mathbf{B})(\mathbf{C})}}(t=0,z^{(\mathbf{D})})=0,
\end{equation}
\begin{equation}
\varpi^{(\mathbf{A})}(t=0,z^{(\mathbf{D})})=0 ,
\end{equation}
and
\begin{equation}
\omega^{(\mathbf{A})}(t=0,z^{(\mathbf{D})})=dtz^{(\mathbf{A})} .
\end{equation}
Consider, at an n+1-dimensional manifold, a coordinate system given
by $(t,z^{(\mathbf{A})})$. For each value of t we have a
hyper-surface, where $dt=0$ on each of them. We are interested in
the hyper-surface with $t=1$, where we verify the following equality
\begin{equation}
\omega^{(\mathbf{A})}(t=1,z)=\varpi^{(\mathbf{A})}(t=1,z).
\end{equation}
From the above results \cite{3},
\begin{equation}
\frac{\partial^2(A_{(\mathbf{A}){(\mathbf{C})(\mathbf{D})}})}{\partial(t^2)}=
tz^{(\mathbf{B})}R_{(\mathbf{A})(\mathbf{B})(\mathbf{C})(\mathbf{D})}+
z^{(\mathbf{L})}z^{(\mathbf{M})}R_{(\mathbf{A})(\mathbf{L})(\mathbf{M})(\mathbf{N})}
A_{(\mathbf{P}){(\mathbf{C})(\mathbf{D})}}\eta^{(\mathbf{N})(\mathbf{P})}.
\end{equation}
Using the curvature symmetries we have the following solution
\begin{equation}
A_{(\mathbf{A}){(\mathbf{C})(\mathbf{D})}}+A_{(\mathbf{A}){(\mathbf{D})(\mathbf{C})}}=0,
\end{equation}
that is true for all t.
\newline
Then,
\begin{equation}
A_{(\mathbf{A}){(\mathbf{C})(\mathbf{D})}}=-A_{(\mathbf{A}){(\mathbf{D})(\mathbf{C})}},
\end{equation}
so that,  we can rewrite $(3.9)$ as
\begin{equation}
\varpi^{(\mathbf{A})}= tdz^{(\mathbf{A})}+
\frac{1}{2}A^{({A})_{(\mathbf{B})(\mathbf{C})}}(z^{(\mathbf{B})}dz^{(\mathbf{C})}-z^{(\mathbf{C})}dz^{(\mathbf{B})}).
\end{equation}
Let us  define
\begin{equation}
A_{(\mathbf{A}){(\mathbf{C})(\mathbf{D})}}=z^{(\mathbf{B})}B_{(\mathbf{A}){(\mathbf{B})(\mathbf{C})(\mathbf{D})}}.
\end{equation}
The following result is obtained by substituting $(3.19)$ in
$(3.15)$,
\begin{equation}
\frac{\partial^2(B_{(\mathbf{A}){(\mathbf{B})(\mathbf{C})(\mathbf{D})}})}{\partial(t^2)}=
tR_{(\mathbf{A})(\mathbf{B})(\mathbf{C})(\mathbf{D})}+
z^{(\mathbf{L})}z^{(\mathbf{M})}R_{(\mathbf{A})(\mathbf{B})(\mathbf{L})(\mathbf{N})}
B_{(\mathbf{P}){(\mathbf{M})(\mathbf{C})(\mathbf{D})}}\eta^{(\mathbf{N})(\mathbf{P})}.
\end{equation}
Using the curvature symmetries we obtain the solution
\begin{equation}
B_{(\mathbf{A}){(\mathbf{B})(\mathbf{C})(\mathbf{D})}}+B_{(\mathbf{B}){(\mathbf{A})(\mathbf{C})(\mathbf{D})}}=const.,
\end{equation}
for all t.
\newline
We can obtain
\begin{equation}
B_{(\mathbf{A}){(\mathbf{B})(\mathbf{C})(\mathbf{D})}}+B_{(\mathbf{B}){(\mathbf{A})(\mathbf{C})(\mathbf{D})}}=0.
\end{equation}
In the following, for future use, we present the line element on the
hyper-surface
\begin{equation}
 ds'^2=\eta_{(\mathbf{A})(\mathbf{B})}\varpi^{(\mathbf{A})}\varpi^{(\mathbf{B})}.
\end{equation}
We conclude that
$B_{(\mathbf{A}){(\mathbf{B})(\mathbf{C})(\mathbf{D})}} $ has the
same symmetries of the Riemann curvature tensor
\begin{equation}
B_{(\mathbf{A}){(\mathbf{B})(\mathbf{C})(\mathbf{D})}}=-B_{(\mathbf{B}){(\mathbf{A})(\mathbf{C})(\mathbf{D})}}=
-B_{(\mathbf{A}){(\mathbf{B})(\mathbf{D})(\mathbf{C})}}.
\end{equation}
 Using $(3.19)$ and $(3.24)$ we have
\begin{eqnarray}
\nonumber A_{(\mathbf{A}){(\mathbf{C})(\mathbf{D})}}dz^{(\mathbf{A})}z^{(\mathbf{C})}dz^{(\mathbf{D})}=\\
\nonumber +\frac{1}{4}B_{(\mathbf{A}){(\mathbf{B})(\mathbf{C})(\mathbf{D})}}.\\
\nonumber .(z^{(\mathbf{B})}dz^{(\mathbf{A})}-z^{(\mathbf{A})}dz^{(\mathbf{B})}).\\
\nonumber .(z^{(\mathbf{C})}dz^{(\mathbf{D})}-z^{(\mathbf{D})}dz^{(\mathbf{C})}).\\
\end{eqnarray}
By direct use of $(3.23)$, $(3.25)$, and $(3.18)$ we have
\begin{eqnarray}
\nonumber ds'^2=t^2\eta_{(\mathbf{A})(\mathbf{B})}dz^{(\mathbf{A})}dz^{(\mathbf{B})}+\\
 \nonumber+\frac{1}{2}\{\frac{1}{2}t\epsilon_{(\mathbf{B})}B_{(\mathbf{A}){(\mathbf{B})(\mathbf{C})(\mathbf{D})}}+\\
 \nonumber+\eta^{(\mathbf{M})(\mathbf{N})}A_{(\mathbf{M}){(\mathbf{B})(\mathbf{A})}}A_{(\mathbf{N}){(\mathbf{C})(\mathbf{D})}}\}.\\
\nonumber.(z^{(\mathbf{B})}dz^{(\mathbf{A})}-z^{(\mathbf{A})}dz^{(\mathbf{B})})(z^{(\mathbf{C})}dz^{(\mathbf{D})}-z^{(\mathbf{D})}dz^{(\mathbf{C})}).\\
\end{eqnarray}
The line elements of the manifold and the hyper-surface are equal at
$ t=1 $, where  $u^{\Lambda}=v^{\Lambda} $,
\begin{equation}
ds^2=ds'^2,
\end{equation}
and
\begin{eqnarray}
\nonumber ds^2=\eta_{(\mathbf{A})(\mathbf{B})}dz^{(\mathbf{A})}dz^{(\mathbf{B})}+\\
\nonumber+\frac{1}{2}\{\frac{1}{2}\epsilon_{(\mathbf{B})}B_{(\mathbf{A}){(\mathbf{B})(\mathbf{C})(\mathbf{D})}}+\\
\nonumber+\eta^{(\mathbf{M})(\mathbf{N})}A_{(\mathbf{M}){(\mathbf{B})(\mathbf{A})}}A_{(\mathbf{N}){(\mathbf{C})(\mathbf{D})}}\}.\\
\nonumber.(z^{(\mathbf{B})}dz^{(\mathbf{A})}-z^{(\mathbf{A})}dz^{(\mathbf{B})})(z^{(\mathbf{C})}dz^{(\mathbf{D})}-z^{(\mathbf{D})}dz^{(\mathbf{C})}).\\
\end{eqnarray}
It can also be written in the form
\newpage
\begin{eqnarray}
\nonumber[1-\frac{1}{2}[\frac{1}{2}\epsilon_{(\mathbf{B})}B_{(\mathbf{A}){(\mathbf{B})(\mathbf{C})(\mathbf{D})}}+\\
\nonumber+\eta^{(\mathbf{M})(\mathbf{N})}A_{(\mathbf{M}){(\mathbf{B})(\mathbf{A})}}A_{(\mathbf{N}){(\mathbf{C})(\mathbf{D})}}].\\\nonumber.(z^{(\mathbf{B})}\frac{dz^{(\mathbf{A})}}{ds}-z^{(\mathbf{A})}\frac{dz^{(\mathbf{B})}}{ds})(z^{(\mathbf{C})}\frac{dz^{(\mathbf{D})}}{ds}-z^{(\mathbf{D})}\frac{dz^{(\mathbf{C})}}{ds})]ds^2\\
\nonumber =\eta_{(\mathbf{A})(\mathbf{B})}dz^{(\mathbf{A})}dz^{(\mathbf{B})}.\\
\end{eqnarray}
We now define the function
\begin{eqnarray}
L^{(\mathbf{A})(\mathbf{B})}=(z^{(\mathbf{A})}\frac{dz^{(\mathbf{B})}}{ds}-z^{(\mathbf{B})}\frac{dz^{(\mathbf{A})}}{ds}),
\end{eqnarray}
which is the classical angular momentum  of a free particle.
\newline
The line element $(3.29)$ can assume the following form
\begin{eqnarray}
 \nonumber\{1+\frac{1}{2}[\frac{1}{2}(\epsilon_{(\mathbf{B})}B_{(\mathbf{A}){(\mathbf{B})(\mathbf{C})(\mathbf{D})}}+\\
 \nonumber+\eta^{(\mathbf{M})(\mathbf{N})}A_{(\mathbf{M}){(\mathbf{B})(\mathbf{A})}}A_{(\mathbf{N}){(\mathbf{C})(\mathbf{D})}}].\\
\nonumber.(L^{\mathbf{A})(\mathbf{B})}L^{\mathbf{C})(\mathbf{D})})\}ds^2\\
 \nonumber =(\eta_{(\mathbf{A})(\mathbf{B})}dz^{(\mathbf{A})}dz^{(\mathbf{B})}.\\
\end{eqnarray}
We now define the function
\begin{eqnarray}
\nonumber \exp(-2\sigma)=\{1+\frac{1}{2}[\frac{1}{2}(\epsilon_{(\mathbf{B})}B_{(\mathbf{A}){(\mathbf{B})(\mathbf{C})(\mathbf{D})}}\\
 \nonumber +\eta^{(\mathbf{M})(\mathbf{N})}A_{(\mathbf{M}){(\mathbf{B})(\mathbf{A})}}A_{(\mathbf{N})){(\mathbf{C})(\mathbf{D})}})].\\
\nonumber .L^{(\mathbf{A})(\mathbf{B})}L^{(\mathbf{C})(\mathbf{D})}\},\\
\end{eqnarray}
so that, the line element assumes the form
\begin{equation}
ds^2=\exp(2\sigma)\eta_{(\mathbf{A})(\mathbf{B})}dz^{(\mathbf{A})}dz^{(\mathbf{B})}.
\end{equation}
It is conformal to an n-dimensional flat manifold, as well as to an
n-dimensional manifold of constant curvature, when, in normal
coordinates, they are well-behaved in the origin and in its
neighborhood. In this paper, for general relativity, we have  $n=4$.
In this case there is a time $\tau$ and (3.33) can be written in the
particular form, as follows
\newpage
\begin{eqnarray}
\nonumber ds^2=\eta_{(\mathbf{a})(\mathbf{b})}dz^{(\mathbf{a})}dz^{(\mathbf{b})}+\\
 \nonumber+\{\eta_{(\mathbf{0})(\mathbf{0})}+\frac{1}{2}[\frac{1}{2}\epsilon_{(\mathbf{B})}B_{(\mathbf{A}){(\mathbf{B})(\mathbf{C})(\mathbf{D})}}+\\
 \nonumber+\eta^{(\mathbf{M})(\mathbf{N})}A_{(\mathbf{M}){(\mathbf{B})(\mathbf{A})}}A_{(\mathbf{N}){(\mathbf{C})(\mathbf{D})}}].\\
\nonumber.(z^{(\mathbf{B})}\frac{dz^{(\mathbf{A})}}{d\tau}-z^{(\mathbf{A})}\frac{dz^{(\mathbf{B})}}{d\tau})(z^{(\mathbf{C})}\frac{dz^{(\mathbf{D})}}{d\tau}-z^{(\mathbf{D})}\frac{dz^{(\mathbf{C})}}{d\tau})\}d\tau^2,\\
\end{eqnarray}
where$(a),(b)\neq 0 $.
 \newline
 Defining
\begin{eqnarray}
d\rho^2=\{\eta_{(\mathbf{0})(\mathbf{0})}+\frac{1}{2}[\frac{1}{2}\epsilon_{(\mathbf{B})}B_{(\mathbf{A}){(\mathbf{B})(\mathbf{C})(\mathbf{D})}}+\\
 \nonumber+\eta^{(\mathbf{M})(\mathbf{N})}A_{(\mathbf{M}){(\mathbf{B})(\mathbf{A})}}A_{(\mathbf{N}){(\mathbf{C})(\mathbf{D})}}].\\
\nonumber.(z^{(\mathbf{B})}\frac{dz^{(\mathbf{A})}}{d\tau}-z^{(\mathbf{A})}\frac{dz^{(\mathbf{B})}}{d\tau})(z^{(\mathbf{C})}\frac{dz^{(\mathbf{D})}}{d\tau}-z^{(\mathbf{D})}\frac{dz^{(\mathbf{C})}}{d\tau})\}d\tau^2,\\
\end{eqnarray}
then, $(3.34)$ can be rewritten as
\begin{equation}
ds^2=d\rho^2+\eta_{(\mathbf{a})(\mathbf{b})}dz^{(\mathbf{a})}dz^{(\mathbf{b})},
\end{equation}
\newline
where in the coordinates $\rho$ and $z^{(\mathbf{a})}$ $(3.37)$ is a
flat metric. Note that this is true where the  Riemannian normal
coordinates are well-behaved in the origin and in its neighborhood.
\newpage
\renewcommand{\theequation}{\thesection.\arabic{equation}}
\section{\bf Mapping Among Manifolds}
\setcounter{equation}{0}
 $         $
In this section we present a short review of a modification of the
Hamiltonian formalism, obtained from a rupture with the symplectic
hypothesis \cite{4}. For an extension of the modified Hamiltonian
formalism, see \cite{5}. Mapping among manifolds are possible in
this modified formalism.
\newline
It is well-known that in the Hamiltonian formalism the Hamilton
equations and the Poisson brackets will be conserved only by a
canonical or symplectic transformation. In the modified-Hamiltonian
formalism only Hamilton equations will be conserved, in the sense
that they will be transformed into other Hamilton equations by a
non-canonical or non-symplectic transformation, and the Poisson
brackets will not be invariant. We now build a modified Hamiltonian
formalism. Consider a time-dependent Hamiltonian $H({\tau})$ where
${\tau}$ is an affine parameter, in this case, the proper-time of
the particle. Let us define 2n variables that will be called
${\xi}^j$ with index j running from 1 to 2n so that we have
${\xi}^j$ $\in$
$({\xi}^1,\ldots,{\xi}^n,{\xi}^{n+1},\ldots,{\xi}^{2n})$ =$(
{q}^1,\ldots,{q}^n,{p}^1,\ldots,{p}^n)$ where ${q}^j$ and ${p}^j$
are coordinates and momenta, respectively. We now define the
Hamiltonian by
\begin{equation}
 H({\tau})=\frac{1}{2}H_{ij}{\xi}^i{\xi}^j,
\end{equation}
where $H_{ij}$ is a  symmetric matrix. We impose that the
Hamiltonian obeys the Hamilton equation
\begin{equation}
\frac{d{\xi}^i}{d\tau }={J}^{ik}\frac{\partial{H}}{\partial{\xi}^k }
.
\end{equation}
The equation $(4.2)$ introduces the symplectic  J, given by
\begin{equation}
\left(%
\begin{array}{cc}
  O & I \\
  -I & O \\
\end{array}%
\right)
\end{equation}
where O and I are the $n \textbf{x}n$ zero and identity matrices,
respectively. We now make a linear transformation from ${\xi}^j$ to
${\eta}^j$ given by
\begin{equation}
  {\eta}^j={{T}^j}_k{\xi}^k,
\end{equation}
where ${{T}^j}_k$ is  a non-symplectic matrix, and the new
Hamiltonian is given by
\begin{equation}
 Q=\frac{1}{2}C_{ij}{\eta}^i{\eta}^j,
\end{equation}
where $C_{ij}$ is a  symmetric matrix. The matrices H, C, and T obey
the following system
\begin{equation}
 \frac{d{{T}^i}_j}{d\tau}+\frac{d{t}}{d\tau}{{T}^i}_k{J}^{kl}X_{lj}=J^{im}Y_{ml}{{T}^l}_j,
\end{equation}
where $2X_{lj}=\frac{\partial{H_{ij}}}{\partial{\xi}^l
}\xi^{i}+2H_{lj}$ and
$2Y_{ml}=\frac{\partial{C_{il}}}{\partial{\eta}^m
}\eta^{i}+2C_{ml},$ t and $\tau$ are the proper-times of the
particle in two different manifolds. We note that $(4.6)$ is a first
order linear differential equation system in ${{T}^i}_k ,$ and it is
the response to what we looked for because the non-linearity in the
Hamilton equations were  transferred to their coefficients. Consider
$\frac{d{t}}{d\tau}X_{lj}=Z_{lj}$ and write $(4.6)$ in the matrix
form
\begin{equation}
 \frac{d{T}}{d\tau}+TJZ=JYT,
\end{equation}
where T, Z and Y are  $2n \textbf{x}2n$ matrices as
\begin{equation}
\left(%
\begin{array}{cc}
  T_{1} & T_{2} \\
  T_{3} & T_{4} \\
\end{array}%
\right)
\end{equation}
with similar expressions for Z and Y. Let us  write $(4.7)$ as
follows
\begin{equation}
 \dot{T_1}=Y_{3}T_{1}+Y_{4}T_{3}+T_{2}Z_{1}-T_{1}Z_{3},
\end{equation}
\begin{equation}
 \dot{T_2}=Y_{3}T_{2}+Y_{4}T_{4}+T_{2}Z_{2}-T_{1}Z_{4},
\end{equation}
\begin{equation}
 \dot{T_3}=-Y_{1}T_{1}-Y_{2}T_{3}+T_{4}Z_{1}-T_{3}Z_{3},
\end{equation}
\begin{equation}
 \dot{T_4}=-Y_{1}T_{2}-Y_{2}T_{4}+T_{4}Z_{2}-T_{3}Z_{4}.
\end{equation}
Now consider
\begin{equation}
 \dot{S_1}=Y_{3}S_{1}+Y_{4}S_{3},
\end{equation}
\begin{equation}
 \dot{S_2}=Y_{3}S_{2}+Y_{4}S_{4},
\end{equation}
\begin{equation}
 \dot{S_3}=-Y_{1}S_{1}-Y_{2}S_{3},
\end{equation}
\begin{equation}
 \dot{S_4}=-Y_{1}S_{2}-Y_{2}S_{4},
\end{equation}
\newpage
and
\begin{equation}
 \dot{R_1}=R_{2}Z_{1}-R_{1}Z_{3},
\end{equation}
\begin{equation}
 \dot{R_2}=R_{2}Z_{2}-R_{1}Z_{4},
\end{equation}
\begin{equation}
 \dot{R_3}=R_{4}Z_{1}-R_{3}Z_{3},
\end{equation}
\begin{equation}
 \dot{R_4}=R_{4}Z_{2}-R_{3}Z_{4}.
\end{equation}
From the theory of  first order differential equation systems [4],
it is well-known that the system $(4.13)-(4.20)$ has a solution in
the region where $Z_{lj}$ and $Y_{ml}$ are continuous functions. In
this case, the solution for $(4.6)$ or $(4.7)$ is given by
\begin{equation}
 {T_1}=(S_{1}a+S_{2}b)R_{1}+(S_{1}d+S_{2}c)R_{3},
\end{equation}
\begin{equation}
{T_2}=(S_{1}a+S_{2}b)R_{2}+(S_{1}d+S_{2}c)R_{4},
\end{equation}
\begin{equation}
{T_3}=(S_{3}a+S_{4}b)R_{1}+(S_{3}d+S_{4}c)R_{3},
\end{equation}
\begin{equation}
{T_4}=(S_{3}a+S_{4}b)R_{2}+(S_{3}d+S_{4}c)R_{4},
\end{equation}
where a,b,c and d are constant $n \textbf{x}n$ matrices, and
substituting $(4.21)-(4.24)$ in $(4.4)$ we will have completed the
mapping among manifolds.
\newline
Although this approach is much more general than the one in Section
$3$, we will need the latter, because mapping, embedding, and other
operations are easier in  Riemannian normal  coordinates.
\newpage
\renewcommand{\theequation}{\thesection.\arabic{equation}}
\section{\bf The Mapping }
\setcounter{equation}{0}
 $         $
In this section we will build a mapping between two metrics. These
two metrics will be $(2.8)$ and another one representing the
ordinary matter. When a Riemannian normal coordinate is well-behaved
in the origin and in its neighborhood, the metric associated with
the ordinary matter can be put in the form $(3.33)$ as follows
\begin{equation}
ds^2=\exp(2\sigma)\eta_{(\mathbf{A})(\mathbf{B})}dz^{(\mathbf{A})}dz^{(\mathbf{B})}.
\end{equation}
Note that the metric (2.8) can be put in the form (5.1) as follows
\begin{equation}
T=\int{e^{-\frac{1}{2}[-2\in mb(\sqrt{\frac{A}{3}})t-\frac{m^{2}}{3}
t^{2}]}}dt,
\end{equation}
\begin{equation}
\Psi=e^{\frac{1}{2}[-2\in mb(\sqrt{\frac{A}{3}})t-\frac{m^{2}}{3}
t^{2}]},
\end{equation}
so that (2.8) assume the following form
\begin{equation}
ds^{2}={\Psi^{2}}[dT^{2}-d\sigma^{2}].
\end{equation}
Using the results of section $4$ we will build the mapping between
$(5.1)$ and $(5.4)$. Note, from $(4.4)$, the dependence of the
coordinates and momenta associated with the ordinary matter from the
coordinates and momenta associated with (2.8).
\newpage
\renewcommand{\theequation}{\thesection.\arabic{equation}}
\section{\bf The Embedding }
\setcounter{equation}{0}
 $         $
The embedding of only one metric  in a flat space  is well known,
\cite{8}. For the embedding of (2.8) and a classical metric in a
$6$-dimensional flat space we proceed as follows.
\newline
Let us write the metric associated with the ordinary matter in the
form $(3.33) $
\begin{equation}
ds^2=\exp(2\sigma)\eta_{(\mathbf{A})(\mathbf{B})}dz^{(\mathbf{A})}dz^{(\mathbf{B})}.
\end{equation}
For the ordinary matter metric $(6.1)$, we define the following
 transformation of coordinates
\begin{equation}
y^{(\mathbf{A})}=\exp(\sigma)z^{(\mathbf{A})},
\end{equation}
with $ (A)=(1,2,3,4),$
\begin{equation}
y^{5}=\exp(\sigma)(\eta_{(\mathbf{A})(\mathbf{B})}z^{(\mathbf{A})}z^{(\mathbf{B})}-\frac{1}{4}),
\end{equation}
and,
\begin{equation}
y^{6}=\exp(\sigma)(\eta_{(\mathbf{A})(\mathbf{B})}z^{(\mathbf{A})}z^{(\mathbf{B})}+\frac{1}{4}).
\end{equation}
It is easy to verify that
\begin{equation}
\eta_{\mathbf{A}\mathbf{B}}y^{\mathbf{A}}y^{\mathbf{B}}=0,
\end{equation}
where,
\begin{equation}
\eta_{\mathbf{A}\mathbf{B}}=(\eta_{(\mathbf{A})(\mathbf{B})},\eta_{\mathbf{(5),}\mathbf{(5)}},\eta_{\mathbf{(6),}\mathbf{(6)}}),
\end{equation}
with,
\begin{equation}
\eta_{\mathbf{(5),}\mathbf{(5)}}=1,
\end{equation}
and,
\begin{equation}
\eta_{\mathbf{(6),}\mathbf{(6)}=-1}.
\end{equation}
By a simple calculation we can verify that the line elements are
given by
\begin{equation}
ds^2=\exp(2\sigma)\eta_{(\mathbf{A})(\mathbf{B})}dz^{(\mathbf{A})}dz^{(\mathbf{B})}=
\eta_{\mathbf{A}\mathbf{B}}dy^{\mathbf{A}}dy^{\mathbf{B}}.
\end{equation}
\newpage
The equation (6.5) is a hyper-cone in the (6)-dimensional flat
manifold. The metric (6.1) was embedded in the hyper-cone (6.5) of
the (6)-dimensional flat manifold. In this paper $n=4$, so that the
hyper-cone represents a region in a $6$-dimensional flat manifold,
as the light-cone in the Minkowski's spacetime, although, with a
different physical meaning.
\newline
For the metric $(2.8)$ we have the following transformation of
coordinates

\begin{equation}
y'^{1}={\Psi}.T,
\end{equation}
\begin{equation}
y'^{\mathbf{i+1}}=\Psi.x^{\mathbf{i}},
\end{equation}
where $i=(1,2,3)$,
\begin{equation}
y'^{5}={\Psi}(\eta_{{\alpha}{\beta}}x^{\alpha}x^{\beta}-\frac{1}{4}),
\end{equation}
and,
\begin{equation}
y'^{6}={\Psi}(\eta_{{\alpha}{\beta}}x^{\alpha}x^{\beta}+\frac{1}{4}).
\end{equation}
It is easy to verify that
\begin{equation}
\eta_{{\alpha}{\beta}}y'^{\alpha}y'^{\beta}=0,
\end{equation}
where,
\begin{equation}
\eta_{{\alpha}{\beta}}=(\eta_{{1}{1}},...,\eta_{\mathbf{5}\mathbf{5}},\eta_{\mathbf{6}\mathbf{6}}),
\end{equation}
with, $\eta_{\mathbf{1}\mathbf{1}}=\eta_{\mathbf{5}\mathbf{5}}=1$,
and
$\eta_{\mathbf{2}\mathbf{2}}=\eta_{\mathbf{3}\mathbf{3}}=\eta_{\mathbf{4}\mathbf{4}}=\eta_{\mathbf{6}\mathbf{6}}=-1.$
\newline
By a simple calculation we can verify that the line elements are
given by
\begin{equation}
ds^2=dt^{2}-d\sigma ^{2}e^{[-2\in
mb(\sqrt{\frac{A}{3}})t-\frac{m^{2}}{3} t^{2}]}=
\eta_{{\alpha}{\beta}}dy'^{\alpha}dy'^{\beta}.
\end{equation}
Using the results of section $4$ we can build the mapping between
$(6.9)$ and $(6.16)$. Note, from $(4.4)$, the dependence of the
coordinates and the momenta associated with the ordinary matter from
coordinates and momenta associated with $(2.8)$. This is more
evident from the system $(4.21)-(4.24)$.  Note that the metrics are
in different regions of the $6$-dimensional flat manifold, and the
coordinates in different regions of the hyper-cone.
\newpage
\renewcommand{\theequation}{\thesection.\arabic{equation}}
\section{\bf Mapping Among Hyper-Vectors }
\setcounter{equation}{0}
 $         $
In this section we present the two $6$-dimensional hyper-vectors,
normal to the manifold $(2.8)$. They can define the directions of
the matter-energy flows between the two manifolds embedded in the
hyper-cone. Some conventions and results of \cite{9} will be used.
\newline
Rewrite $(6.10)$, $(6.11)$,  $(6.12)$ and $(6.13)$
\begin{equation}
y'^{\mathbf{1}}=\Psi.T,
\end{equation}
\begin{equation}
y'^{\mathbf{i+1}}=\Psi.x^{\mathbf{i}},
\end{equation}
where $i=(1,2,3)$, and
\begin{equation}
y'^{\mathbf{5}}=\Psi[T^{2}-(x^{\mathbf{1}})^{2}-(x^{\mathbf{2}})^{2}-(x^{\mathbf{3}})^{2}-\frac{1}{4}],
\end{equation}
\begin{equation}
y'^{\mathbf{6}}=\Psi[T^{2}-(x^{\mathbf{1}})^{2}-(x^{\mathbf{2}})^{2}-(x^{\mathbf{3}})^{2}+\frac{1}{4}].
\end{equation}
Let us call $\eta'{_{(a)}{^{\alpha}}}$ the two $6$-dimensional
hyper-vectors, normal to the manifold (2.8), where $(a)=(1,2)$, and
$\alpha=(1,2,3,4,5,6)$. By a simple but long calculation we obtain
\cite{9},
\begin{equation}
\eta'{_{(1)}{^1}}=-r[\frac{a}{2}-b't]+[\frac{1}{2r}-\frac{r}{2}(\frac{a}{2}-b't)^{2}]y'^{1},
\end{equation}
\begin{equation}
 \eta'{_{(1)}{^i}}=[\frac{1}{2r}-\frac{r}{2}(\frac{a}{2}-b't)^{2}]y'^{i+1},
\end{equation}
\begin{equation}
 \eta'{_{(1)}{^k}}=[\frac{1}{2r}-\frac{r}{2}(\frac{a}{2}-b't)^{2}]y'^{k}-2r[(\frac{a}{2}-b't)T+\dot{T}],
\end{equation}
and
\begin{equation}
\eta'{_{(2)}{^1}}=r[\frac{a}{2}-b't]+[\frac{1}{2r}+\frac{r}{2}(\frac{a}{2}-b't)^{2}]y'^{1},
\end{equation}
\begin{equation}
 \eta'{_{(2)}{^i}}=[\frac{1}{2r}+\frac{r}{2}(\frac{a}{2}-b't)^{2}]y'^{i+1},
\end{equation}
\begin{equation}
 \eta'{_{(2)}{^k}}=[\frac{1}{2r}+\frac{r}{2}(\frac{a}{2}-b't)^{2}]y'^{k}-2r[(\frac{a}{2}-b't)T+\dot{T}],
\end{equation}
where $i=(1,2,3)$, $k=(5,6)$, $a=-2\in mb(\sqrt{\frac{A}{3}})$,
$b'=\frac{m^{2}}{3}$, $\dot{T}=\frac{dT}{dt}$, and $r=r(t)$ is given
by
\begin{equation}
 (\frac{a}{2}-b't)^{2}=\exp{(b'r^{2})}[c+2b'\int{\frac{1}{r}\exp{(-b'r^{2})}dr}],
\end{equation}
and $c$ is an integration constant.
\newline
From simple calculation we can verify the following conditions,
\begin{equation}
\eta_{\mathbf{A}\mathbf{B}}\eta'{_{(1)}{^{\mathbf{A}}}}\eta'{_{(1)}{^{\mathbf{B}}}}=1,
\end{equation}
\begin{equation}
\eta_{\mathbf{A}\mathbf{B}}\eta'{_{(2)}{^{\mathbf{A}}}}\eta'{_{(2)}{^{\mathbf{B}}}}=-1.
\end{equation}
\newline
Note that the two $6$-dimensional hyper-vectors $(7.12)$ and
$(7.13)$, normal to the manifold $(2.8)$, live in the
$6$-dimensional flat manifold, while $6$-dimensional hyper-vectors,
as $(6.5)$ and $(6.14)$, live in the hyper-cone. It is possible that
the hyper-vectors associated with the ordinary matter manifold,
represented by $(6.9)$, could be integrable. In this case, from
Section $4$, we can build a Hamiltonian function to each
hyper-vector and the mapping between them. Note, from (4.4), the
coordinate and momentum dependence between the ordinary matter
metric and (2.8), where $\eta{_{(a)}{^{\alpha}}}$ and
$\eta'{_{(a)}{^{\alpha}}}$ are coordinates associated with the
ordinary matter metric and (2.8), respectively. The hyper-vectors
can define the directions of the matter-energy flows between the two
embedded manifolds. From (4.4) we conclude that the mapping connects
these flows.
\newpage
\section{Concluding Remarks}
  $              $
Physicists and astronomers are convinced  of the dark matter and
dark energy existence. There is a great interest in it and many
researchers have concentrated their efforts in verifying and solving
such hypothesis with a view point different from the one presented
in this paper, [9], [10], [11], [12], [13], [14]. This motivates the
emergence of new methods, new formalisms. In this paper we have used
an extended Cartan's approach, where pseudo-Riemannian metrics are
conformal to flat manifolds when Riemannian normal coordinates are
well-behaved in the origin and in its neighborhood. We also
presented a modified Hamiltonian formalism, where the symplectic
hypothesis was abandoned. This formalism allows us to make the
mapping  among geometric objects as metrics and hyper-vectors, for
instance.
\newline
It is important to pay attention to the fact that a Riemannian
normal transformation and its inverse are well-behaved in the region
where geodesics  are not mixed. Points where geodesics close or mix
are known as conjugate points of Jacobi's fields. Jacobi's fields
can be used for this purpose, \cite{3}.  After we put two metrics in
the conformal flat form, we accomplish the map between them,
obtaining (4.4), where coordinates and momenta will be connected,
exerting a mutual gravitational influence. We made the same
procedure for the hyper-vectors. An interesting application could be
a mapping between (2.8) and the Schwarzschild's metric. In general
(4.4) is invertible, and it enables an analysis of the gravitational
influence between the metrics. Another interesting investigation
would be a mapping between (2.8) and a galaxy. Also an interesting
investigation would be a mapping between two hyper-vectors. It is
possible that the hyper-vectors associated with the ordinary matter
manifold could be integrable, obtaining the two $6$-dimensional
hyper-vectors, normal to the ordinary matter manifold. In this case,
from section $4$, we can build a Hamiltonian function for each
hyper-vector and the mapping between them. Note, from (4.4), the
coordinate and momentum dependence between the ordinary matter
metric and (2.8), where $\eta{_{(a)}{^{\mathbf{A}}}}$ and
$\eta'{_{(a)}{^{\mathbf{A}}}}$ are coordinates associated with the
ordinary matter metric and (2.8), respectively. The hyper-vectors
can define the directions of the matter-energy flows between the two
embedded manifolds. From (4.4) we conclude that the mapping connects
the flows.  We do not know if those mapping represent a physical
reality, as the matter-energy flows between the two embedded
manifolds. The formalism presented in Section $4$ allows us to make
mapping among different metrics, different geometric objects, etc.
However, our choice of a 6-dimensional flat manifold was motivated
by the possibility that normal hyper-vectors could define the
direction of matter-energy flows between two embedded manifolds, if
these flows there exist.


\begin{thebibliography}{99}
\bibitem{1} A.C.V.V.de Siqueira, arXiv: 0101012v1[gr-qc]
\bibitem{2} A.C.V.V.de Siqueira, arXiv: 10096193v1[gr-qc]
\bibitem{3} A.C.V.V.de Siqueira, arXiv: 10062868v1[math-ph]
\bibitem{4} A.C.V.V.de Siqueira, arXiv: 08022299v1[math-ph]
\bibitem{5} A.C.V.V.de Siqueira, arXiv: 08031124v2[math-ph]
\bibitem{6}  Ya. B. Zel'dovich, Sov. Phys. {USPEKHI, vol. 11}, 1968.
\bibitem{7} T. Padmanabhan, Phys. Rept. 380, 235 (2003).
\bibitem{8} Joe Rosen, Rev. Mod. Phys. 37, 204(1965).
\bibitem{9}L.P.Eisenhart,{\bf Riemannian Geometry}(Princeton University Press,1997)
\newline
There is a small mistake in the equation (47.4) which spreads along
some Sections. It is important to correct it. In (47.4), replace
$\epsilon_{(\sigma)}$ by one. This implies a small change in other
equations, as (47.5) and (47.6). In other words,
$\epsilon_{(\sigma)}$ will have to appear in (47.5) and (47.6), for
instance.
\bibitem{10}P.J.E.Peebles, arXiv: 0207347v2[astro-ph]
\bibitem{11} T. Padmanabhan,  arXiv: 08021798v1[gr-qc]
\bibitem{12}T. Padmanabhan, arXiv: 08072356v1[gr-qc]
\bibitem{13}G.R.Farrar and P.J.E.Peebles, arXiv: 0307316v2[astro-ph]
\bibitem{14}S.Micheletti,E.Abdalla and B.Wang,  arXiv: 09020318v4[gr-qc]
\end{thebibliography}
\end{document}